# Theory of Epitaxial Growth of Borophene on Layered Electride: Thermodynamic Stability and Kinetic Pathway

Xiaojuan Ni,[1] Huaqing Huang,[1] Kyung-Hwan Jin,[1] Zhengfei Wang[2] and Feng Liu[1,3,*]

[1]Department of Materials Science and Engineering,
University of Utah, Salt Lake City, Utah 84112, USA

[2]Hefei National Laboratory for Physical Sciences at the Microscale,
CAS Key Laboratory of Strongly-Coupled Quantum Matter Physics,
University of Science and Technology of China, Hefei, Anhui 230026, China

[3]Collaborative Innovation Center of Quantum Matter, Beijing 100084, China

**Abstract**

Based on first-principles calculations, we propose that a layered electride $Mg_2O$ can serve as an effective substrate, in place of metal substrate, to grow honeycomb borophene (h-borophene). We first confirm the thermodynamic stability of h-B@$Mg_2O$ heterostructure by energetics analysis and dynamic stability by absence of imaginary frequency in phonon spectra. Then, kinetically, we identify the atomistic pathways for a preferred 2D growth mode over 3D growth, and reveal a growth transition from 2D compact islands to h-borophene at ~13-atom cluster size, indicating the feasibility of epitaxial growth of h-borophene on $Mg_2O$. The h-borophene is found to be stabilized by nearly one electron transfer from $Mg_2O$ to h-borophene based on the Bader charge analysis. Moreover, the intrinsic band structure of the free-standing h-borophene is only weakly perturbed by the $Mg_2O$ substrate, a significant advantage over metal substrate. We envision that layered electrides provide an attractive family of substrates for epitaxial growth of a range of 2D materials that can otherwise only be grown on undesirable metal substrates.

Two-dimensional (2D) materials have drawn intensive attention for both fundamental interests and potential applications [1–4]. While high quality small-size 2D materials, such as graphene [5,6] and transition metal dichalcogenides [7,8], may be prepared by mechanical exfoliation, it is highly desirable to grow 2D materials for large size and mass quantity to further our study and realize their potential applications. However, many 2D materials can only be grown on metal substrates, which is unwanted because the ample metallic states will inevitably destroy the intrinsic, especially non-metallic properties of 2D materials. In this Letter, using honeycomb borophene (h-borophene, a monolayer boron sheet) as a prototypical example, we demonstrate that layered electrides provide a new class of substrates, in place of metal, to effectively grow 2D materials with significant advantages.

Boron is one of the most versatile elements, forming various 3D bulk polymorphs [9–13]. Similarly, different 2D borophene have been theoretically proposed [14–18] and experimentally realized [19–21]. Most excitingly, one recent experiment succeeded in growing h-borophene on Al(111) surface [21]. Analogous to graphene, a free-standing h-borophene will host massless Dirac fermions and exhibit fascinating electronic properties. Furthermore, in the well-known high-$T_c$ superconductor, $MgB_2$, boron also possesses a planar h-borophene structure, which is primarily responsible for the superconductivity [22,23]. A recent study also showed $MgB_2$ to be a topological semimetal [24]. Thus, synthesis of h-borophene may not only provide new opportunities for 2D high-$T_c$ superconductivity but also topological superconductivity. However, the growth of h-borophene on metal substrate has some drawbacks. It is unlikely to exfoliate h-borophene away from the metal substrate. The high conductivity of metal will mask the intrinsic transport properties of h-borophene due to high electron density of states of metal near the Fermi level [25]. Therefore, searching for an insulating or weak metal substrate to grow h-borophene is of great interest.

Since boron has three valence electrons, the electron deficiency makes its honeycomb lattice energetically unstable. Thus, neither semiconductors nor insulators can serve as good substrates to grow h-borophene. Instead, we have identified layered electride, a "weak metal", to be an effective substrate to grow h-borophene.

Electrides are crystals with cavity-trapped electrons acting as anions. They were first synthesized by Dye in 1983 [26], in the form of 0D and 1D confined electrons [27–29]. Interestingly, a layered electride, di-calcium nitride ($Ca_2N$), was experimentally reported in 2013 to possess delocalized "2D electron gas (2DEG)" sandwiched between the cationic layers [30]. $Ca_2N$ was then studied as an efficient electron donor for hydrogenation in alkynes and alkenes [31]. And several carbides [32–36], nitrides [32,34–38], and oxides [32,33,36] have since been predicted as layered electrides. We hypothesize that the presence of 2DEG on the layered electride surface provides an effective means to compensate the electron deficiency of h-borophene. In addition, the layered structure of electride can facilitate the exfoliation or growth of h-borophene on monolayer or thin films of electrides to minimize the substrate influence. Finally, as a weak metal, the significantly reduced density of states from the electride at the Fermi level helps to retain the most, if not all, intrinsic electronic and transport properties of h-borophene.

To test the above hypotheses, we investigate both thermodynamic stability of h-B@$Mg_2O$ and kinetic pathways of boron growth on $Mg_2O$ surface using first-principles calculations. The dynamic stabilities of $Mg_2O$ and heterostructure of h-B@$Mg_2O$ are confirmed by the absence of imaginary frequency in phonon spectra. The calculated electron localization function (ELF) indicates the existence of interlayer 2DFEG in $Mg_2O$. Nearly one electron transfers from $Mg_2O$ substrate to each boron atom to stabilize h-borophene based on the Bader charge analysis. We identify the atomistic kinetic pathways for boron growth on $Mg_2O$ surface with a preferred 2D growth mode over 3D mode. A critical growth transition from compact-triangular structure to hexagonal-ring structure is revealed with increasing boron cluster size. All these findings point to the feasibility of epitaxial growth of h-borophene on layered electride.

The layered electride $Mg_2O$ is chosen to have an appropriate lattice constant as the prototypical substrate to grow h-borophene. Fig. 1(a) and (b) depict the crystal structures of monolayer and bulk $Mg_2O$, similar to $Ca_2N$ [30] and $Y_2C$ [39,40]. Bulk $Mg_2O$ is in the R-3m space group with lattice constant of 3.01 Å. The building block of $Mg_2O$ consists of a triple layer where O atoms are sandwiched by top and bottom Mg layers. We calculated the phonon dispersion for monolayer and bulk $Mg_2O$ to check the dynamic stability, as shown in Fig. 1(d) and (e), respectively. No imaginary frequency exists in the spectra. The free-standing h-borophene possesses a planar honeycomb structure with lattice constant of 2.93 Å, comparable to that of $Mg_2O$. Fig. 1(c) depicts the top view of the heterostructure of h-B@$Mg_2O$. The topmost Mg atoms in $Mg_2O$ are located below the hollow positions of h-borophene, which is the most stable configuration. The phonon dispersion of h-B@$Mg_2O$ in Fig. 1(f) shows no imaginary frequency, which confirms the dynamical stability of h-B@$Mg_2O$.

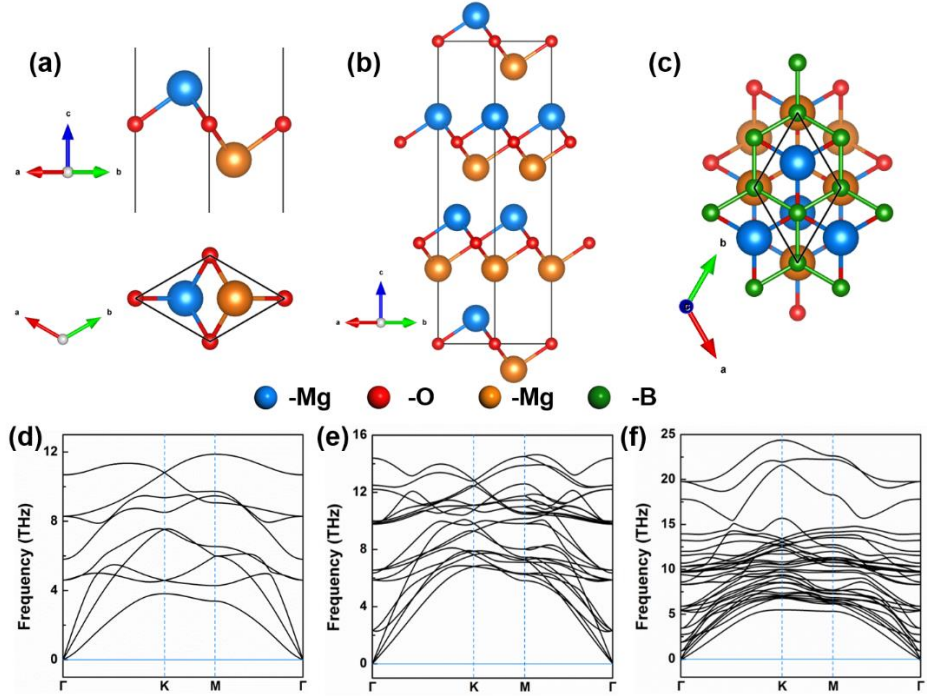

**Fig. 1.** Crystal structures: (a) Top and side views of monolayer $Mg_2O$; (b) Side view of bulk $Mg_2O$; (c) Top view of h-B@$Mg_2O$. The black lines indicate the unit cell. The phonon spectra of (d) monolayer $Mg_2O$, (e) bulk $Mg_2O$, and (f) h-B@$Mg_2O$.

It is crucial to characterize $Mg_2O$ as an electride in a chemical formula $[Mg_2O]^{m+} \cdot me^-$, where m is the number of electrons, and confirm the existence of 2DEG. The ground-state band structure of monolayer $Mg_2O$ is shown in Fig. 2(a) without spin polarization. The two parabolic bands arising from Mg-$s$ and O-$p$ orbitals indicate the existence of 2DEG. This is also confirmed by the plots of partial charge density in Fig. S1(a) to (c) in Supplemental Materials [41]. The degree of electron delocalization can be quantitatively described by ELF. On both surfaces of monolayer $Mg_2O$, a delocalized ELF feature is clearly shown in Fig. 2(b). Also, for $Mg_2O$ monolayer with one and two valence electrons removed, the delocalized ELF features on both surfaces gradually disappear [Fig. 2(c) and (d)]. This indicates that electron transfer away from $Mg_2O$ will first vacate the 2DFEG states, suggesting that the 2DEG can be used as an effective electron donor to compensate the electron deficiency in h-borophene upon growth. The ELF analysis also indicates that $Mg_2O$ is a layered electride with a formula of $[Mg_2O]^{2+} \cdot 2e^-$. In bulk, the anionic electrons locate within the interlayer space between the $[Mg_2O]^{2+}$ layers [see, e.g., Fig. S1(d) and (e)] [41]. Since there are two boron atoms in one unit cell of h-B@$Mg_2O$, one electron is expected to transfer from $Mg_2O$ substrate to each boron atom to stabilize h-borophene.

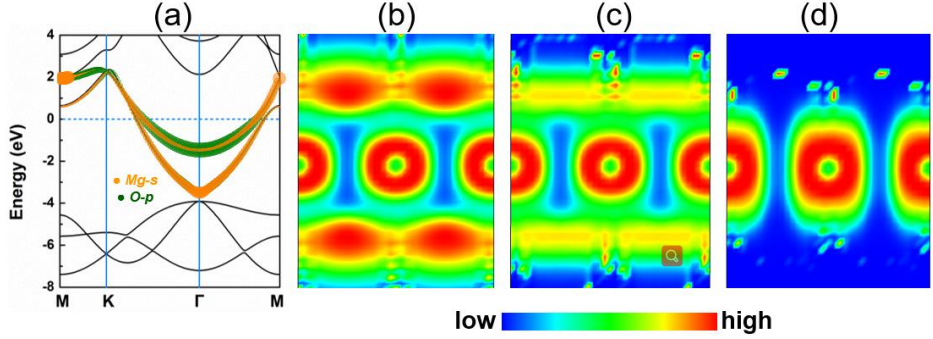

**Fig. 2.** (a) Band structure of monolayer Mg$_2$O. The orange and green dots represent the contributions from Mg-*s* orbital and O-*p* orbital, respectively. (b) Mg$_2$O, (c) [Mg$_2$O]$^+$, and (d) [Mg$_2$O]$^{2+}$ ELF maps with a color bar from low density to high density.

Previous theoretical studies have already indicated that free-standing h-borophene is unstable [42]. The substrate used to grow h-borophene must play a crucial role in stabilizing the honeycomb lattice. To better understand the stabilization mechanism, we calculate the formation energy of h-borophene on different substrates and perform the Bader charge analysis [43]. The formation energy is defined as:

$$E_{form} = \frac{E_{tot} - E_{sub} - N \times E_B}{N}, \qquad (1)$$

where $E_{tot}$ is the total energy of h-borophene on the substate, $E_{sub}$ is the energy of the substrate, $E_B$ is the energy of the isolated boron atom, and *N* is the number of boron atoms [44]. Different from previously calculations [21,45], we take the energy of isolated boron atom as $E_B$ instead of the cohesive energy of bulk α-B$_{12}$ because atomic boron source prepared by e-beam evaporator under ultrahigh vacuum condition is used during growth [19–21]. The calculated formation energy for h-B@Mg$_2$O and h-B@Al(111) are -6.28 and -6.35 eV/atom, respectively. According to the Bader charge analysis, about 0.6 electrons on average are transferred from Mg$_2$O to each boron atom, which is about the same as that (0.7 electrons) in h-B@Al(111) [21]. Apparently, there is no significant difference in formation energy and electron transfer between h-B@Mg$_2$O and h-B@Al(111) [21], which indicates that the layered electride Mg$_2$O could potentially replace metal Al to grow h-borophene.

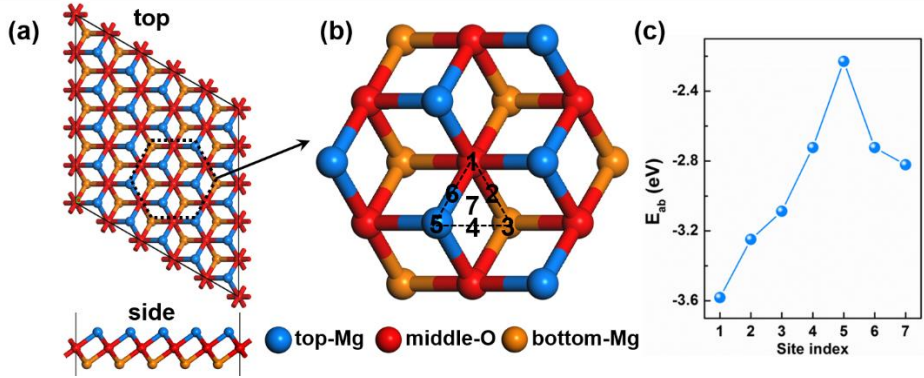

**Fig. 3.** (a) Top and side views of a 5×5 supercell of monolayer $Mg_2O$ substrate. (b) Illustration of absorption Site **1** to **7** for boron adatom on $Mg_2O$ by zooming in the dashed hexagon region in (a). (c) Absorption energy for Site **1** to **7**.

Next, we establish the preferred growth mode and identify the atomic pathways for boron growth on $Mg_2O$ substrate, by investigating initial stage of growth up to 13 atoms. Although growth of borophene on metal substrates [45–47] have been studied previously, fundamental mechanisms of borophene growth on layered electride are unknown. We adopted a 5×5 monolayer $Mg_2O$ as the substrate [Fig. 3(a)], which is confirmed to be sufficiently large to avoid inter-cluster interaction for up to 13-atom clusters.

We first determine the preferred adsorption site for boron adatom. Fig. 3(b) shows seven different absorption sites on $Mg_2O$ surface, as labeled **1** to **7**, and Fig. 3(c) shows the corresponding adsorption energy. The most stable one is Site **1** [top of middle-O atom, see Fig. 4(a)], followed by Site **2** (center of bridge$_{1-3}$) and **3** (top of bottom-Mg atom). A boron adatom is unstable on other four positions, especially Site **5** (above top-Mg atom). Thus, the boron adatom is kinetically hindered from climbing onto Site **5**. Also, the distance between Site **1** and **2** is shorter than the B-B bond length, so that these two sites cannot be occupied simultaneously owing to Coulomb repulsion. Hence, we tend to use Site **1** and **3** to construct the geometries for boron clusters at the initial growth stage.

After structural relaxation, the preferred stable configurations of small boron clusters (N = 1 to 5) are depicted in Fig. 4(a) to (e), respectively. As illustrated in Fig. 4(b), the most stable positions for a boron dimer are Site **1** and **3**, which is also expected from the adatom absorption energy. For a larger cluster, such as a trimer, a chain-type structure occupying Site **1** and **3** is most stable, as shown in Fig. 4(c). Adding another adatom to the chain-trimer, it prefers to diffuse to the periphery to form a 2D rhomboid-shaped tetramer [Fig. 4(d)], with two atoms locating in Site **1** and one locating in Site **3**. By continuously adding adatom, the pentamer [Fig. 4(e)], hexamer [Fig. S4(a)], heptamer [Fig. S4(b)] and octamer [Fig. S4(c)] [41] will form in sequence, and they all have the 2D structures as the most stable configurations. In particular, all the preferred 2D configurations have the adatoms occupying exclusively Site **1** and **3** to form a compact-triangular structure.

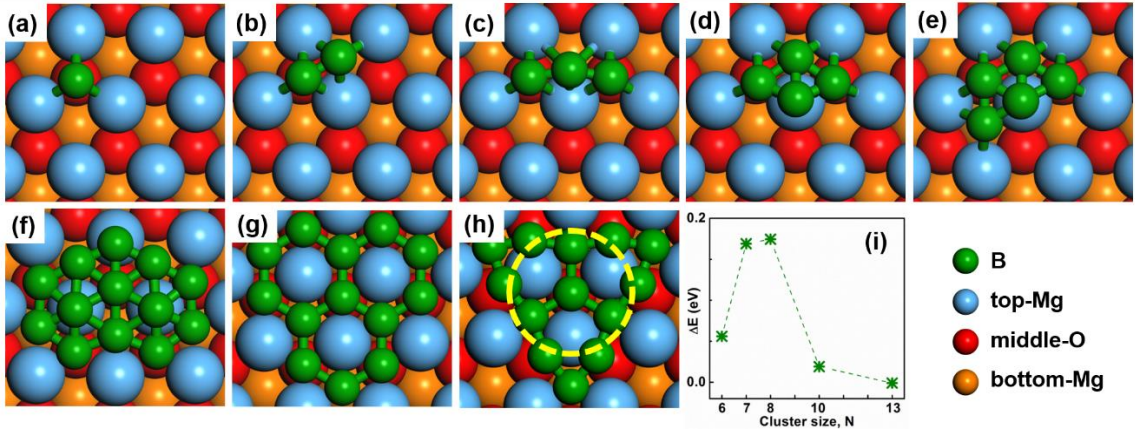

**Fig. 4.** Preferred 2D configurations of boron clusters on $Mg_2O$ surface: (a) adatom, (b) dimer, (c) trimer, (d) tetramer, (e) pentamer., and (f) compact cluster with N = 13. (g) Initial and (h) final structures of N = 13 cluster with 6-atom rings. (i) Energy difference between ring and compact configurations for boron clusters (N = 6, 7, 8, 10 and 13).

As a comparison, we also constructed the initial cluster configurations with a hexagonal ring, which is the desired building block for h-borophene, as shown for N = 13 in Fig. 4(g). For N < 13, such ring structures [e.g., N = 6, 7, 8, and 10 in Fig. S4(i) to (l)] are unstable, and will spontaneously transform into a nearly equilateral-triangle shape [Fig. S4(e) to (h)], which are metastable compared to the compact islands [Fig. S4(a) to (d)] [41]. Importantly, however, for N = 13, the atom in the middle of the equilateral-triangle structure has three nearest neighbors, which assume already the geometry of h-borophene, as indicated by the yellow-dashed circle in Fig. 4(h). By evaluating the relative stability between the ring and compact configurations, we found that their energy differences are 0.056, 0.169, 0.174, 0.019, and -0.001 eV/atom, respectively, as depicted in Fig. 4(i). The significant energy difference for N = 7 and 8 is due to the unsaturated atoms on the periphery [see Fig. S4(f) and (g)] [41]. Most importantly, however, the ring cluster in Fig. 4(h) is more stable than the compact one [Fig. 4(f)]. This indicates that with the increasing cluster size, a growth transition from the equilateral-triangle shape to h-borophene will occur at N = 13. We further confirmed this at the infinite 2D limit of a periodic structure. We constructed the initial configuration of borophene with 6-atom equilateral-triangle cluster to begin with, and it spontaneously transforms into the h-borophene, as illustrated in Fig. S5 [41]. The above results imply that the electride $Mg_2O$ substrate facilitates not only the 2D growth mode of planar island but also a transition to the ultimate h-borophene sheet with the increasing coverage of boron atoms during growth.

Finally, beyond establishing the growth of h-borophene on $Mg_2O$ substrate, we further compare the monolayer $Mg_2O$ versus four-layer Al(111) for growing h-borophene. Most notably, the electronic states of h-borophene grown on $Mg_2O$ are much less perturbed than those grown on Al(111), as evidenced by much less metallic states coming from the former than the latter, as shown in Fig. S6 [41]. This signifies a distinctive advantage of layered electride substrate over conventional metal substrate for growing h-borophene.

In conclusion, we propose for the first time that a layered electride $Mg_2O$ can serve as an effective substrate to grow h-borophene. Both thermodynamic stability of h-B@$Mg_2O$ and kinetic pathways of boron growth have been investigated using first-principles calculations. The 2DEG in $Mg_2O$ surface will compensate the electron deficiency in h-borophene to stabilize the honeycomb structure and facilitate the 2D growth mode. Our findings open a new avenue for epitaxial growth of h-borophene using layered electride substrates with notable advantages over metal substrates, which will stimulate immediate theoretical and experimental interests. Broadly, the layered electrides may be used as effective substrates for epitaxial growth of various 2D materials that can otherwise only be grown on undesirable metal substrates.

This work is supported by U.S. Department of Energy-Basic Energy Sciences (Grant No. DE-FG02-04ER46148). We acknowledge DOE-NERSC and CHPC at the University of Utah for providing the computing resources.

*Corresponding author.

Email: fliu@eng.utah.edu